\documentclass[twocolumn,prl,aps,showpacs]{revtex4}

\usepackage[dvips]{graphicx}
\usepackage{amsmath}
\usepackage{amsfonts}
\usepackage{amssymb}%
\input epsf

\def\be{\begin{equation}}

\def\ee{\end{equation}}

\def \be {\begin{equation}}
\def \ee {\end{equation}}
\def \bea {\begin{eqnarray}}
\def \eea {\end{eqnarray}}

\begin{document}


\begin{titlepage}

\hspace*{\fill}\parbox[t]{4cm}{
UW/PT 05-04\\
hep-ph/0503110\\
\today}

\bibliographystyle{apsrev}
\preprint{UW/PT 05-04}
\title {Using Charge Asymmetries to Measure Single Top Quark Production at the LHC}
\author {Matthew T. Bowen}
\affiliation {Department of Physics, PO Box 351560, University of Washington, 
  Seattle, WA 98195, U.S.A.\\
  {\tt [mtb6@u.washington.edu]}}

\begin{abstract}
Electroweak production of single top quarks is an as-yet-unverified prediction of the 
Standard model, potentially sensitive to new physics.  Two of the single top quark
productions channels have significant charge asymmetries at the LHC, while the much larger
background from $t\bar{t}$ is nearly charge-symmetric.  This can be used to 
reduce systematic uncertainties and make precision measurements of single top 
quark production. 

\end{abstract}
\maketitle
\end{titlepage}

\section{Electroweak Production of Single Top Quarks}

The electroweak production of single top quarks at hadron colliders \cite{dicus} is an 
important prediction of the Standard Model which remains to be verified. 
Limits from Run I at the Tevatron have been published \cite{runI}, and new limits
from Run II are just now emerging \cite{runII}.
Measuring single top quark production
is important because it equates to the first direct measurement of the CKM matrix element $V_{tb}$.
In addition, anomalous top quark couplings
and a variety of proposed new physics models affect the three single-top-quark 
production
modes in largely orthogonal ways.  Thus, independent measurements of all three
channels provide a possible window to new physics \cite{tait}, as well as for 
some discrimination between possible models.

\begin{figure}
[ptbhptbh]
\begin{center}
\includegraphics[
height=2.2cm,
width=6.6cm
]%
{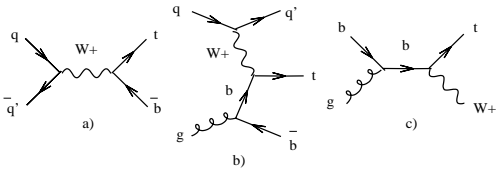}
\caption{Feynman diagrams for a) $tb$, b) $tbq$, and c) $tW$ single top production.}%
\label{tb_tbq}%
\end{center}
\end{figure}

One single-top-quark production mode, 
``$tb$'', occurs through diagrams such as Figure \ref{tb_tbq}a,
where an $s$-channel $W$ boson creates a top-bottom quark pair.  A second, the so-called
``$W$-gluon fusion'' or ``$tbq$'', mode occurs through diagrams such as Figure \ref{tb_tbq}b,
where a $t$-channel $W$ boson fuses with a bottom quark from gluon splitting to 
form a top.  The ``$tW$'' production mode \cite{twtait} occurs when a bottom
quark and a gluon scatter to create a top quark and a real $W$ boson, as in Figure \ref{tb_tbq}c.
This last mode, which is smaller than $tbq$, but larger than $tb$,
is cleanly removed by  
the method outlined below, and will not be discussed at length.

The detector signature for $tb$ and $tbq$ single-top production comes primarily from 
the decay products
of the top quark: one lepton, missing energy transverse
to the beam (MET) from a neutrino, and one $b$ quark --- all at high $p_{T}$.  
In addition, the $tb$ channel has a second
large-$p_T$ $b$ quark, while the $tbq$ channel has both a low-$p_{T}$ $b$ quark from gluon 
splitting, which is rarely visible,
and a moderately high-$p_{T}$ light quark.  After showering and hadronization, 
short-distance quarks and gluons lead to ``jets''
of energy in detector calorimeters.
To reduce backgrounds,
at least one jet in single top event selection samples is
usually required to be ``$b$-tagged'' (identified as containing
a metastable hadron, probably with a $b$ quark constituent.)

At the LHC, the largest background to the $tb$ and $tbq$ signal is $t\bar{t}$ 
pair production \cite{ssw}, where one or both of the top quarks decays leptonically, but only one 
lepton is identified.
These events also have high-$p_T$ jets from $b$ quarks, and typically one or 
two other high-$p_T$ jets from light quarks.   
A second background is $W$ boson production with associated jets.
For the present paper, this background will be divided into two sets:  $Wb\bar{b}$, where
the $b$-tags come from $b$ quarks produced perturbatively in the hard scattering; and 
$Wjj$, where the $b$-tags come from mistagging of charm, gluon, and light quark jets, and 
also the fragmentation of short-distance gluons to $b\bar{b}$ pairs. 
A third background comes from pure QCD processes, where a lepton is either 
misidentified or comes from the decay of a heavy-quark meson.  The MET for these
QCD events comes from jet energy fluctuations in the calorimeter.  This background
is not estimated here, but is argued to be small for the method described below.

The definitive signal-to-background study of single top quark production 
at the LHC has remained reference \cite{ssw} since its publication.  
The size of the $t\bar{t}$ background led the 
authors of \cite{ssw} to study a ``jet veto'',
wherein events with more than 2 jets were rejected.  As $t\bar{t}$ typically has 
more than 2 jets per event, the jet veto removed greater than $90\%$ of this
background.  For the single $b$-tag sample, this technique was sufficient to study the $tbq$
channel. For the double $b$-tag sample, however, the jet veto could not
sufficiently reduce $t\bar{t}$ to permit a study of either the $tbq$ or $tb$ 
channels. 

\subsection{Charge Asymmetries}

Fortunately, single top quark production is a very unusual
process, with other features
that can be used to separate it from backgrounds.  At the Tevatron, single-top production
has significantly larger parity asymmetries than its Standard Model backgrounds
which can be used to help isolate the single top signal \cite{BES}.
This is a result of the parity-asymmetric
$p\bar{p}$ initial state at the Tevatron, versus the charge-asymmetric $pp$ 
initial state at the LHC.
At the LHC, single-top production has large charge asymmetries
which can be used to separate it from backgrounds.  
For both the $tb$ and $tbq$ channels, roughly $60-70\%$ more top quarks are produced at the 
LHC than anti-top quarks, and this difference
is preserved in the charge of the leptons from the top quark decays.

Define $N_{+}$ ($N_{-}$) to be the number of events with one high-$p_T$
positively- (negatively-) charged lepton.  Then define $N_{total}=N_++N_-$,
$\Delta = N_+ - N_-$, and a charge asymmetry $A_C=\Delta / N_{total}$.
For the cuts defined in table \ref{cuts1}, the $tbq$ channel has a charge asymmetry of 
order $26\%$ and is the largest contributor
to $\Delta$.  The $tb$ channel has a charge asymmetry of order $20\%$, but, because
of its much smaller cross-section, does not contribute as significantly to $\Delta$.

To first approximation, the $t\bar{t}$ background has a charge asymmetry of zero.
But $N_+$ and $N_-$ are functions of both the event kinematic distributions
and the acceptance regions of the detectors.  At leading-order (LO), the $t$ and $\bar{t}$
distributions are identical, but at NLO 
there is a preference for $\bar{t}$ quarks to be more central than $t$ quarks \cite{kuhn},
so more $t\bar{t}$ events with leptons ($N_-$) will be detected than
with anti-leptons ($N_+$).  However, the magnitude of this asymmetry 
is estimated below to be smaller than $0.05\%$. 

QCD backgrounds should have smaller charge asymmetries than $t\bar{t}$.  For example, the
$b\bar{b}$ background has smaller
NLO corrections to its distributions than $t\bar{t}$ \cite{kuhn}.  In addition, neutral
$B$-meson mixing dilutes the charge correlation between the $b$ quarks and leptons,
further reducing the contribution to $\Delta$.  And to the extent 
that fake leptons from other QCD processes are charge symmetric, these also should 
not contribute significantly to $\Delta$.  

The $W$+jets background, on the other hand, is more complicated.  
The sub-channels for this background do have significant asymmetries of varying
magnitudes and signs.  This will be discussed further below, but the size
and uncertainty of this background's contribution to $\Delta$ do not 
appear to be prohibitively large.

Finally, the 
$tW$ production mode is charge-symmetric because the $b$ and $\bar{b}$ parton
distribution functions (PDFs) are believed to be charge-symmetric.  
There are also no known 
higher-order corrections
to the kinematic distributions that would introduce a measurable charge asymmetry for this mode, 
and any future corrections are anticipated to be small. 
Thus, considering charge asymmetries
provides a clean way of separating the $tbq$ and $tb$ modes from the $tW$ mode.

\section{Signal and Background Study}

For the study below, MadEvent \cite{Maltoni:2002qb} is used to generate both signal and background
event samples with the CTEQ5 \cite{cteq} PDF set.  
Single top $tb$ events are generated with 
$\mu_{F}$=$\mu_{R}$=$m_t$=175 GeV (where $\mu_{F}$ and $\mu_R$ are the factorization and 
renormalization scales) and normalized to  
$6.55$ ($4.07$) pb \cite{harris} for $t\bar{b}$ ($\bar{t}b$) production. 
The $t\bar{b}q$ event set is generated using the 
same scales as \cite{ssw}: $\mu_F^2=-q_W^2$ (where $q_W$ is the t-channel W boson momentum) 
for the initial state quark PDF, and 
$\mu_F^2=\mu_R^2=m_b^2+p_{T\bar{b}}^2$ for the initial state gluon PDF. 

At NLO, the $tbq$ channel receives large contributions when the $b$ or $\bar{b}$ quark
from gluon-splitting is at low $p_T$.  In order to reflect this enhancement in a lowest-order (LO)
simulation, reference \cite{ssw} normalized the $tbq$ sample 
differentially depending on
the $p_T$ of the $b$ or $\bar{b}$
from gluon-splitting (from here on, the $b$ or $\bar{b}$ is referred to simply as ``the $b$ quark'').  
In \cite{ssw}, the cross-section with the $b$ quark $p_T$ above 20 GeV 
was determined at
LO to be 81 pb, and the cross-section with the $b$ quark $p_T$ below 20 GeV
was normalized to 164 pb to match the total
NLO cross-section of 245 pb \cite{sswnlo}.  

References \cite{ssw} and \cite{sswnlo} were not concerned with separating
the $t\bar{b}q$ and $\bar{t}bq$ channels, so only the sum of the two was quoted.  They are
separated here by computing the ratio of the LO $t\bar{b}q$ and $\bar{t}bq$ cross-sections with
the $b$ quark $p_T$ above 20 GeV using Madevent \cite{Maltoni:2002qb}, and multiplying 
by 81 pb \cite{ssw}.  
For the $b$ quark $p_T$ above 20 GeV, this leads to 52 pb for $t\bar{b}q$ and 
29 pb for $\bar{t}bq$.  The $t\bar{b}q$ sample with the $b$ quark $p_T$ below 20 GeV is normalized to
104 pb, so that when added to 52 pb, the total $t\bar{b}q$ cross-section is equal to the 
total NLO rate of 156 pb \cite{harris}.  Likewise, 
the $\bar{t}bq$ sample with the $b$ quark $p_T$ below 20 GeV
is normalized to 62 pb to match the total NLO rate of 91 pb \cite{harris}.  Note that 
the quoted total $tbq$ cross-sections in \cite{sswnlo} and \cite{harris} differ by $1\%$,
which is negligible for the current analysis.    

The $t\bar{t}$ sample is generated with $\mu_R=\mu_F=m_t$ and normalized to 
873 pb \cite{Kidonakis:2003qe}.
The $Wb\bar{b}$ and $Wjj$ sets are generated with
$\mu_{F}$=$\mu_{R}$=$M_W$ and normalized with K-factors = 2.35 and 0.87 \cite{Campbell:2003hd}, 
respectively.  These
K-factors are taken from the study done in \cite{Campbell:2003hd}, under the assumption 
that they are the same for the slightly different cuts used here.
The $W$+jets background is only simulated for $W$-plus-two jets, since the uncertainty
in this prediction is probably as large as the $W$-plus-three jets contribution 
is likely to be.

Partons are mapped to jets by smearing their energies with a gaussian function of width 
$\sigma/E=0.64/\sqrt{E}\oplus 0.036$, from Table 9-1 in \cite{atlastdr}, to simulate showering
and detector response.  Taus
are treated as jets. Leptons are required to be 
separated from jets 
by $\Delta R=0.7$, and jets within
$\Delta R=0.7$ of each other are merged. 

The $b$-tagging parametrizations for jets from bottom quarks, charm quarks, gluons, 
and light quarks are derived from Figure 10-24 of \cite{atlastdr}, which assumes that
jets from $b$ quarks are tagged at a fixed rate of $50\%$. Jets from charm quarks 
are tagged at an efficiency $\epsilon_c(\%)=7.26+123$ $\textrm{GeV}/p_T+0.0044$ $\textrm{GeV}^{-1}\times p_T$, 
gluons at $\epsilon_g(\%)=-14.1+185$ $\textrm{GeV}/p_T+2.9\times \ln{(p_T/\textrm{GeV})}$, and light quarks 
at $\epsilon_q(\%)=-1.5+71$ $\textrm{GeV}/p_T+0.0095\times (p_T/GeV)$. 
For the $W$+jets sample, using the cuts in Table \ref{cuts1}, this is approximately 
equivalent to tagging charm quark, gluon, and light quark jets
at fixed rates of $10\%$, $1.3\%$, and $0.5\%$, respectively.

For this study, events are required to have MET, one and only one charged lepton, and at least
two jets, with one or two of them $b$-tagged. 
Table \ref{cuts1}
lists the cuts used to select events, and Table \ref{nums1} lists the cross-section in
femtobarns (fb) for each channel after branching-ratios and detector
cuts.

\begin{table}[ptb]
\par
\begin{center}%
\begin{tabular}
[c]{|c|c|c|}\hline
Item & $p_{T}$ & $\left\vert \eta\right\vert $ \\\hline
$\ell^{\pm}$ & $\geq20$ GeV & $\leq2.5$\\\hline
MET $\left(  \nu\right)  $ & $\geq20$ GeV & -\\\hline
$b$-tagged jets & $\geq30$ GeV & $\leq2.5$\\\hline
other jets & $\geq30$ GeV & $\leq4.5$\\\hline
\end{tabular}
\end{center}
\caption{Detector cuts used to select events.}%
\label{cuts1}%
\end{table}

\begin{table}[ptb]
\par
\begin{center}%
\begin{tabular}
[c]{|c|c|c|}\hline
Channel & 1 tag & 2 tags \\\hline
$t\overline{b}$    & 277     & 106   \\\hline
$\overline{t}b$    & 178     & 73    \\\hline
$t\overline{b}q$   & 7,410   & 955   \\\hline
$\overline{t}bq$   & 4,300   & 552   \\\hline
$W^+b\overline{b}$ & 1,340   & 531   \\\hline
$W^-b\overline{b}$ & 858     & 351   \\\hline
$W^+jj$            & 12,700  & 79    \\\hline
$W^-jj$            & 10,900  & 76    \\\hline
$t\overline{t}$    & 95,700  & 33,600\\\hline
\end{tabular}
\end{center}
\caption{Cross-sections (in femtobarns) with 1 or 2 $b$-tags required, after 
accounting for leptonic branching 
ratios and detector cuts.  $Wjj$ includes all final
states with charm quarks, gluons, and light quarks.}%
\label{nums1}%
\end{table}

\subsection{The $t\bar{t}$ Asymmetry}

The asymmetry  in $t\bar{t}$ at NLO cannot currently be computed, since there is no NLO
event generator for $t\bar{t}$ which includes the spin correlations for 
top quark decay.  For this paper, an upper bound on the magnitude of the observed 
charge asymmetry for $t\bar{t}$ is estimated using 
Figure 14 in Reference \cite{kuhn}. This figure plots the $t\bar{t}$-pair charge
asymmetry $A_{t\bar{t}}(y)$ for $q\bar{q}$ initial-states as a function of rapidity.
If the detector has acceptance in the rapidity range $-y_o<y<y_o$, the total asymmetry 
observed for $q\bar{q}$ initial states is:
\begin{equation}
A_{q\bar{q}}(y_{o})=\frac{\int_{-y_{o}}^{y_{o}} dy \: \frac{d\sigma_{t\bar{t}}}{dy}(y)\times A_{t\bar{t}}(y)}{\int_{-y_o}^{y_o} dy \: \frac{d\sigma_{t\bar{t}}}{dy}(y)}
\label{asym}
\end{equation}

As $y_{o} \rightarrow \infty$, $A_{q\bar{q}}$
goes to zero, as there are the same number of $t$ and $\bar{t}$ quarks.  For
$y_{o} \rightarrow 0$, $A_{q\bar{q}}$ goes to $-0.25\%$, the value of $A_{t\bar{t}}(y)$ at $y=0$.  Since 
$\frac{d\sigma_{t\bar{t}}}{dy}(y)>0$, $\frac{d\sigma_{t\bar{t}}}{dy}(y)=\frac{d\sigma_{t\bar{t}}}{dy}(-y)$,
$A_{t\bar{t}}(y)=A_{t\bar{t}}(-y)$, and $A_{t\bar{t}}(y)$ is monotonically increasing with $|y|$ 
\cite{kuhn}, any $y_{o}$ will yield an asymmetry $0 \geq A_{q\bar{q}} \geq -0.25\%$.

Since there are no NLO corrections to $gg$ initial
states which introduce charge asymmetries, and the charge-asymmetric 
corrections to $gq$ initial states are small enough to neglect \cite{kuhn}, the overall
bound on the $t\bar{t}$ charge asymmetry can be estimated solely from this bound on $A_{q\bar{q}}$.
At LO, $q\bar{q}$ initial states account for less than $20\%$ of the cross-section at the 
LHC, so the total observed asymmetry $A_C$ for $t\bar{t}$
is estimated to be in the range $0\%>A_C\approx 0.2A_{q\bar{q}}>-0.05\%$.
This bound on the $t\bar{t}$ asymmetry is used here as a surrogate for a bound on the 
observed lepton-antilepton asymmetry, which has not yet been studied. 
Because the $t\bar{t}$ asymmetry is so small, this estimate can be off by 
a factor of two or more without becoming a problem for the measurement.

\subsection{Results}

Table \ref{delta1} shows that the signal-to-background ratio for the single-tag
sample is 1:10 for the total number of events, but around 3:2 for
$\Delta$.  
The contribution from $t\bar{t}$ to $\Delta$ is so small that even large systematic uncertainties
in its prediction are not a problem.
Further, even with only 10 fb$^{-1}$, the measurement of $\Delta$ is not limited by statistics, at
least with the cuts of Table \ref{cuts1}. 

The conclusions for the double-tag sample, shown in Table \ref{delta2}, are much the same, 
except that statistical uncertainties in $t\bar{t}$'s contribution
to $\Delta$ are more significant with only 10 fb$^{-1}$ of data, though still manageable.  
On the other hand, the signal-to-background ratio for $\Delta$ in the double-tag sample is 
2:1, and is thus potentially even less sensitive to systematic errors than the single tag sample.  
With higher statistics,
the double-tag sample could potentially be better than the single-tag sample for a study of  
single-top-quark production.

As this study has been done for a small dataset of 10 fb$^{-1}$, lower detector
efficiencies than assumed here will not be problematic.
In addition, larger $p_T$ cuts on reconstructed objects do not 
decrease the charge asymmetry of the $tbq$ channel.  If conditions at the LHC dictate that all
MET, jet $p_T$, and lepton $p_T$ cuts must be raised to 
50 GeV, the signal will be reduced by roughly 90$\%$ for the single-tag sample. 

\begin{table}[ptb]
\medskip
\par
\begin{center}%
\begin{tabular}
[c]{|c|c|c|c|}\hline
Channel           &  $N_{total}$ & $\Delta$     & $\sqrt{N_{total}}$\\\hline
$tb$              & 4,550        &  990         & 67          \\\hline
$tbq$             & 116,000      &  30,900      & 340        \\\hline
$Wb\bar{b}$       & 21,900       &  4,820       & 150          \\\hline
$Wjj$             & 236,000      &  18,000      & 490        \\\hline
$t\bar{t}$        & 958,000      &  -479        & 980        \\\hline
Total             & 1.34M        &  54,200      & 1,200        \\\hline
\end{tabular}
\end{center}
\caption{Numbers of events with 1 $b$-tag for 10 fb$^{-1}$.  $t\bar{t}$ is assumed to have a $-0.05\%$ charge asymmetry.}%
\label{delta1}%
\end{table}

However, the charge asymmetry method is crucially sensitive to systematic
uncertainties in effects which impact the $N_+$ and $N_-$ samples asymmetrically.  
For example, a systematic uncertainty in the efficiency of positively-charged 
lepton reconstruction of order $0.5\%$, which does not exist for the reconstruction of negatively-charged
leptons, could lead
to an uncertainty in the prediction of $\Delta$ for $t\bar{t}$ which is of the same order
as the signal.
Ensuring that these kinds of uncertainties are sufficiently small will require a detailed 
understanding of the detector response to leptons
and antileptons. 

\begin{table}[ptb]
\medskip
\par
\begin{center}%
\begin{tabular}
[c]{|c|c|c|c|}\hline
Channel           &  $N_{total}$ & $\Delta$  & $\sqrt{N_{total}}$\\\hline
$tb$              & 1,790     &  330         & 42          \\\hline
$tbq$             & 15,100    &  4,030       & 120         \\\hline
$Wb\bar{b}$       & 8,800     &  1,800       & 94          \\\hline
$Wjj$             & 1,550     &   30         & 40          \\\hline
$t\bar{t}$        & 336,000   &  -167        & 580         \\\hline
Total             & 363,000   &  6,020       & 600         \\\hline
\end{tabular}
\end{center}
\caption{Numbers of events with 2 $b$-tags for 10 fb$^{-1}$.  $t\bar{t}$ is assumed to have a $-0.05\%$ charge asymmetry.}%
\label{delta2}%
\end{table}

This opportunity for studying single top quark production at the LHC should be 
contrasted with the current reality at the 
Tevatron.  Currently, there is no clear way to reduce the signal-to-background ratio
to order unity at the Tevatron without problematic systematic and statistical 
uncertainties \cite{BES}.  The LHC, on the other hand, will not be statistics limited, 
and systematic uncertainties can be controlled through the use of charge
asymmetries.  This should allow the LHC to make a precision test of the Standard Model
via single top quark production.

\subsection{Acknowledgements}

The author thanks S. Ellis, A. Haas, and M. Strassler for useful
discussions and suggestions.  This work was supported by U.S. Department
of Energy grant DE-FG02-96ER40956.

\end{document}